**Effect of bombarding steel with Xe+ ions on the surface nanostructure and on pulsed plasma nitriding process**

**Authors:** S. Cucatti, E.A. Ochoa, M. Morales, R. Droppa Jr, J. Garcia, H.C. Pinto, L.F. Zagonel, D. Wisnivesky, C.A. Figueroa, F. Alvarez

# Effect of bombarding steel with Xe[+] ions on the surface nanostructure and on pulsed plasma nitriding process

**S. Cucatti[1,*], E.A. Ochoa[2], M. Morales[1], R. Droppa Jr[3], J. Garcia[4], H.C. Pinto[5], L.F. Zagonel[1], D. Wisnivesky[6], C. A Figueroa[7], and F. Alvarez[1]**

[1] *Instituto de Física "Gleb Wataghin"*, UNICAMP, 13083-859, Campinas, SP, Brazil
[2] *Departamento de Física*, PUCRJ, 22451-900, Rio de Janeiro, Brazil
[3] *Centro de Ciências Naturais e Humanas*, UFABC, 09210-170, Santo André, SP, Brazil
[4] *Sandvik Coromant R&D*, Stockholm, SE-12680, Sweden
[5] *Escola de Engenharia de São Carlos*, EESC-USP, 13566-590, São Carlos, SP, Brazil
[6] *Plasma-LIITS Equipamentos e Processos*, 13083-970, Campinas, SP, Brazil
[7] *CCET, Universidade de Caxias do Sul,* 95070-560, Caxias do Sul-RS, Brazil

[*] Corresponding author. Tel.: +55 19 352 15377; fax +55 19 352 15373; email: scucatti@ifi.unicamp.br

## Abstract

The modification of steel (AISI 316L and AISI 4140) surface morphology and underlying inter-crystalline grains strain due to Xe[+] ion bombardment are reported to affect nitrogen diffusion after a pulsed plasma nitriding process. The ion bombardment induces regular nanometric patterns and increases the roughness of the material surface. The strain induced by the noble gas bombardment is observed in depths which are orders of magnitude larger than the projectiles' stopping distance. The pre-bombarded samples show peculiar microstructures formed in the nitrided layers, modifying the in-depth hardness profile. Unlike the double nitrided layer normally obtained in austenitic stainless steel by pulsed plasma nitriding process, the Xe[+] pre-bombardment treatment leads to a single thick compact layer. In nitrided pre-bombarded AISI 4140 steel, the diffusion zone shows long iron nitride needle-shaped precipitates, while in non-pre-bombarded samples finer precipitates are distributed in the material.

## 1. Introduction

Surface texturing is an important method applied in various technological areas of research. Much of the current interest in surface modifications stems from the possibility of obtaining special optical, tribological, and mechanical properties in a variety of materials [1, 2]. In particular, texturing steels surface by ionic bombardment (atomic attrition) is an interesting route to modify plasma nitriding processes. The efficiency of the process and the final material properties are influenced by modifications of the surface and bulk material such as roughness, defects, and stress created by ion impact. Moreover, a combination of low energy argon ion bombardment and substrate temperature modified diffusion phenomena in semiconductors [3]. Other surface modifications such as mechanical attrition ("shot-penning") generate stress, plastic deformation, defects and roughness. All these changes contribute to modify the kinetic of the surface reactions, shortening the duration of nitriding processes [4,5,6]. Results reported by Abrasonis *et al.* suggest that the effect of bombarding the material with $Ar^+$ ions *after* nitriding processes also has important consequences on nitrogen diffusion in austenitic stainless steel [7]. The observed changes in plasma nitriding processes are attributed to several concomitant causes such as increasing the surface roughness, stress, and creation of lattice defects, altogether improving nitrogen diffusion [8, 9]. Moreover, in the nitriding process, precipitation kinetics lead to possibly complex stress profiles even considering smoothly decreasing nitrogen profiles influencing nitrogen diffusion [10].

In this work we expand the study reported by Ochoa et al. [8, 9] by considering the state of the bombarded surface and the microstrain induced by $Xe^+$ bombardment in stainless and alloyed steels (AISI 316L and AISI 4140) and their influence on the nitrided layer obtained in a long time pulsed plasma nitriding process (20 hours). We remark that in standard pulsed plasma nitriding processes, $Ar^+$ ion bombardment is used in order to clean the material surface by eroding the residual oxide layer hindering nitrogen diffusion. In this paper, we discuss that both the state of the surface (topography) and the microstrain introduced by the ion bombarding should be taken in account in order to better understand the nitriding process. The generated strain is normally associated with diffusion enhancing. Therefore, as the strain induced by the noble gas bombardment extends orders of magnitude beyond the projectiles' stopping length, nitrogen diffusion modifications are expected in pre-bombarded samples. In particular, the $Xe^+$ pre-bombardment treatment of the AISI 316L sample leads to a single compact nitride case. As we shall show below, depending on the crystalline grain orientation, a variety of patterns are formed on the bombarded surface such as ripples, holes, and mounds. Similar considerations are valid for the strain distribution on the bombarded surface. However, we remark that a detailed study of the bombarding effect on each grain´s crystalline orientation is beyond the scope of this paper. Consequently, we shall restrict the study of the ion bombarding effect on the material to a large number of grains, i.e., an *average effect* of the ion bombarding on the nitriding process.

## 2. Experimental

Two series of samples were prepared for the present study: Set I and Set II, consisting of an ensemble of AISI 316L and AISI 4140 samples. Samples from Set I were bombarded by $Xe^+$ and its morphology and strain studied. The samples forming Set II were also bombarded by $Xe^+$ ions and their effect on pulsed plasma nitriding was studied. Details of the sample preparation and characterization procedures follow.

### 2.1. Set I: Sample Preparation.

Rectangular samples of 20x10 mm, 2 mm thick, from the same source of AISI 316L (austenitic stainless steel, nominal composition C: <0.08, Si: <0.5 P: 0.05, S: 0.03, Mn: 1.6, Mo: 2.1, Ni: 12.0, Cr: 17.0, Fe: balanced in wt. %) and AISI 4140 (nominal composition C: 0.4, Si:0.25, P:0.04, S:0.04, Mn: 0,85, Mo: 0.20, Cr: 1, Fe: balanced in w%) were studied. The samples were mirror polished (1400 mesh) using standard metallurgical techniques. The $Xe^+$ ion bombardment was performed at room temperature during 30 minutes using a Kaufman source with 3 cm diameter beam. Details of the ion beam apparatus are described elsewhere [11]. The nominal ion beam energy and current density were fixed at 1 keV and 1.4 mA/cm$^2$, respectively. According to SRIM ion bombarding simulations, the stopping distance of the ions is ~1.1 − 1.8 nm [8, 12]. The background chamber pressure of the deposition system is ~$10^{-5}$ Pa and the working pressure during $Xe^+$ bombardment ~$1.4 \times 10^{-1}$ Pa. For AISI 316L, five impinging ion bombardment angles ($\varphi$ = 0º, 15º, 30º, 45º and 60º) were selected for the study of the morphology after the treatment. In the case of AISI 4140 steel, the effect of ion bombardment under perpendicular impact ($\varphi$ = 0º) was studied. The morphology was analyzed by Scanning Electron Microscopy (FEG-SEM, Quanta 650FEG) and Atomic Force Microscopy (AFM) (Veeco, model Inova), the latter technique used in contact mode.

For the study of the strain produced by ion bombardment, grazing angle x-ray diffraction (GAXRD) measurements were performed in AISI 316L steel for a perpendicularly bombarded sample ($\varphi$ = 0º) and for a reference (only polished) sample. In order to study the in -depth average strain profile, three different X-rays penetration depths were employed by controlling the beam angle of incidence. The effective penetration depth for each x-ray incidence angle was calculated according to Noyan [13], assuming a layer of material contributing with 63% to the total diffracted intensity. The incidence angles and the corresponding calculated penetration depths for a 7.0 keV x-ray beam are shown in Table 1. The diffractograms were measured at the *Brazilian Synchrotron Light Laboratory* (LNLS). A Ge (111) crystal analyzer in front of the scintillation detector permitted to get a high instrument resolution and to make the measurements insensitive to small sample misplacements as well as to possible geometrical aberrations [14].

| Incidence angle, $\alpha$ (º) | Penetration depth ($\mu$m) | | | | |
|---|---|---|---|---|---|
| | (± 0.02 $\mu$m) | | | | |
| | $\gamma$(111) | $\gamma$(200) | $\gamma$(220) | $\gamma$(311) | Average |
| 0.5 | 0.20 | 0.21 | 0.21 | 0.21 | 0.21 |
| 1 | 0.41 | 0.41 | 0.41 | 0.41 | 0.41 |
| 5 | 1.87 | 1.88 | 1.91 | 1.89 | 1.89 |

Table 1: X-ray beam incidence angle ($\alpha$) and the calculated penetration depth of the x-ray ($\lambda$ = 1.7701 Å) for different reflections of Fe (absorption coefficient $\mu$ = 420 cm$^{-1}$).

The Bragg peak widths were analyzed by the Williamson–Hall (W–H) method [15]. Briefly, the W–H method allows a rough estimation of the *average strain* and *average crystallite size* from the integral breadth of diffraction peaks. For this purpose, it is assumed that the total integral breadth $\beta_{hkl}$ of a given diffraction peak (hkl) is a (linear) sum of the contribution from the crystallite size ($\beta_c$) and strain ($\beta_s$) integral breadths, i.e., $\beta_{hkl} = \beta_c + \beta_s$. Here hkl stands for the Miller indexes. These components are given by $\beta_s \cong 2\ \varepsilon \tan \theta$ and $\beta_c \cong K\lambda/L \cos \theta$, respectively, where $\varepsilon$ is the microstrain, K ~ 0.9 is a shape factor proposed by Scherrer assuming cubic-shape grains [16], $\theta$ is the Bragg angle, $\lambda$ and L are the wavelength of the x-ray and the mean linear dimension of the crystallite, respectively. In fact, K depends on the Miller indices of the analyzed reflection, on the crystalline defects, and crystallite shape. However, the assumption of K constant in the W–H analysis is a regular practice since the results do not vary significantly. The instrumental integral breadth was determined by measuring a strain-free LaB$_6$ powder standard (NIST Standard Reference Material 660a) in the same conditions as those in which the samples were measured (Table 2). By adding and rearranging the equation $\beta_{hkl} = \beta_c + \beta_s$, one can write:

$$\beta_{hkl} \cos\theta = 2\varepsilon \sin\theta + K\,\lambda/L = S\sin\theta + Y, \qquad (1)$$

i.e., a linear dependence of $\beta_{hkl}\cos\theta$ on $\sin\theta$ is obtained. Here S=2$\varepsilon$ and Y $=$ K$\lambda$/L stand for the slope and ordinate intercept, respectively. Therefore, by plotting $\beta_{hkl}\cos\theta$ vs. $\sin\theta$ for different (hkl) reflections, the average crystallite sizes L and the microstrain $\varepsilon$ can be estimated from the intercept Y and slope S, respectively [16].

| Penetration depth (µm) | Instrumental integral breadth (rad) | | |
|---|---|---|---|
| (± 0.02 µm) | LaB$_6$ (200) | LaB$_6$ (220) | LaB$_6$ (320) |
| 0.21 | $(11 \pm 1)\times10^{-3}$ | $(13 \pm 1)\times10^{-3}$ | $(20 \pm 1)\times10^{-3}$ |
| 0.41 | $(11 \pm 1)\times10^{-3}$ | $(14 \pm 1)\times10^{-3}$ | $(20 \pm 1)\times10^{-3}$ |
| 1.89 | $(11 \pm 1)\times10^{-3}$ | $(13 \pm 1)\times10^{-3}$ | $(20 \pm 1)\times10^{-3}$ |

Table 2: Instrumental integral breadth obtained from strain-free LaB$_6$ powder standard Bragg reflections.

In order to perform this analysis, the X-ray diffraction peaks were fitted by pseudo-Voigt functions and the total integral breadth calculated using the parameters provided from the fitting according to the equation $[\pi\,m + (1-m)(\pi\,\ln2)^{1/2}]w$ where $m$ is the profile shape factor and $w$ the total peak width [16]. Finally, the integral breadth ($\beta_{hkl}{}^{physical}$) without instrumental contribution was obtained according to the relation $\beta_{hkl}{}^{measured} = \beta_{hkl}{}^{physical} + \beta_{hkl}{}^{instrumental}$.

It is important to remark that the W–H method probes the *micro strain* induced by dislocations and flaws caused by the ion bombarding *inside crystalline grains*, i.e., opposite to a *macro strain* that causes a shift in the Bragg peak position because of the presence of residual macrostresses affecting several grains [16]. Furthermore, the method is traditionally applied to bulk materials in the Bragg-Brentano geometry, where all the probed grains are oriented perpendicularly to the sample surface. In the present case, the W–H method was applied in a grazing incidence configuration as was already done by Tanner et al. [17] in the study of subsurface damage in ceramics. Therefore, as the x-ray beam angle of incidence is kept fixed, the grains probed at different Bragg reflections are oriented in diverse directions relative to the sample surface, introducing some data scattering in the W–H plots. This is particularly true for low x-ray angles of incidence, where the number of grains probed is relatively small (see figures 5 (a) to (c)). Thus, some scatter data can be ascribed to material anisotropy and texture effects [16] and not related to the presence of macrostresses.

## 2.2. Set II: Sample Preparation.

Samples of AISI 316L and AISI 4140 steel (same dimensions and composition of those of Set I) were mirror polished (1400 mesh) using standard metallurgical techniques. Before the pulsed plasma nitriding process, the samples were previously bombarded with $Xe^+$ ions. This pre-bombardment treatments were performed using a unique impinging ion bombardment angle ($\varphi = 0°$, perpendicular bombardment) and the rest of the macroscopic parameters were maintained identical to those described in Section 2.1. In order to minimize experimental ambiguities when comparing the nitriding effect on the pre-bombarded and not bombarded material, the samples were prepared by covering half of the studied surface with a silicon wafer, i.e., the screened part of the sample was not bombarded and the other half part suffered the modifications. Therefore, the two regions, i.e., the pre-bombarded half part and the other not bombarded half (mirror polished) part, undergo identical nitriding conditions from the point of view of the plasma nitriding process. Afterwards, the samples were nitrided during 20 hours in a commercial pulsed plasma vacuum furnace (*Plasma-LIITS, Campinas São Paulo, Brazil*) at 380ºC, ~160 Pa chamber pressure, in a gaseous mixture of nitrogen and hydrogen ( $[N_2]/[H_2+N_2]$=20% ).

The nitrides phases of the nitrided samples were identified using a Bragg-Brentano X-ray diffraction configuration and a monocromatized Cu-K$\alpha$ radiation line (Shimadzu, LAB X-XRD-6000). The cross-section morphology of the studied samples was revealed by etching the material at room temperature with NITAL 5% (95ml ethanol + 5ml $HNO_3$) and Marble (4g $CuSO_4$ in 20ml Cl + 20ml $H_2O$) solutions for the studied AISI 4140 and AISI 316L samples, respectively. The microstructure was studied using a FEG-SEM (Quanta 650FEG). The hardness of the nitrided samples (cross-section) was obtained using a Berkovich diamond tip (NanoTest-100). The load–displacement curves were analyzed using the Oliver and Pharr method [18]. Each experimental point in the hardness curves was obtained averaging 10 measurements. The estimated errors correspond to the standard deviation of the measurements.

## 3. Results and Discussion

### 3.1. Bombarding effect on the material surface

The morphological analysis of the treated samples shows that the $Xe^+$ ion bombarding reveals the crystalline grains and promotes the formation of quite regular patterns, resulting in a nanostructured surface. Figures 1(a), (b), (c), (d) and (e) show examples of sculpted patterns as well as the grains borders obtained in AISI 316L samples bombarded at different selected impinging angles (0º, 15º, 30º, 45º and 60º). We note that at the beginning of the bombarding process the beam angle is fixed relative to the flat polished surface. Afterwards, during the bombarding treatment and due to preferential sputtering, the ion beam is no longer impinging on all the grains with the original fixed angle. Despite some specific peculiarities of the formed patterns for different impinging angles, all the studied samples display similar features, i.e., grooves, ripple periodicity, dunes, terraces, mounds and preferential groove orientation in individual grains. We note that the periodicity of the pattern is quite remarkable while the hills and valleys variations indicate preferential sputtering. The results show that the patterns formed in different grains present diverse directions, i.e., different "k-wave-number" (Figure 1) for the same ions impinging angle. This suggests that some surface accommodation mechanism depends on the crystalline orientation of the each grain, as expected in the Ehrlich–Schwoebel instability model. According to this model, the direction of the pattern periodicity depends, among other things, on the grain crystalline orientation [19,20]. Roughly speaking, the regular pattern stems from essentially two mechanisms inducing surface instability. The first one is related to the surface curvature dependence of the ion *sputtering yield* and the second one is due to the presence of an energy barrier hindering adatoms to diffuse over step edges. The material structure is also important in the process. In a crystalline semiconductor like Si, the stiff covalent directional bonds hinder the atoms' motion and the ripples generally follow the direction of the ion beam. In metals, on the other hand, the ripples generally follow the material crystalline orientation [20].

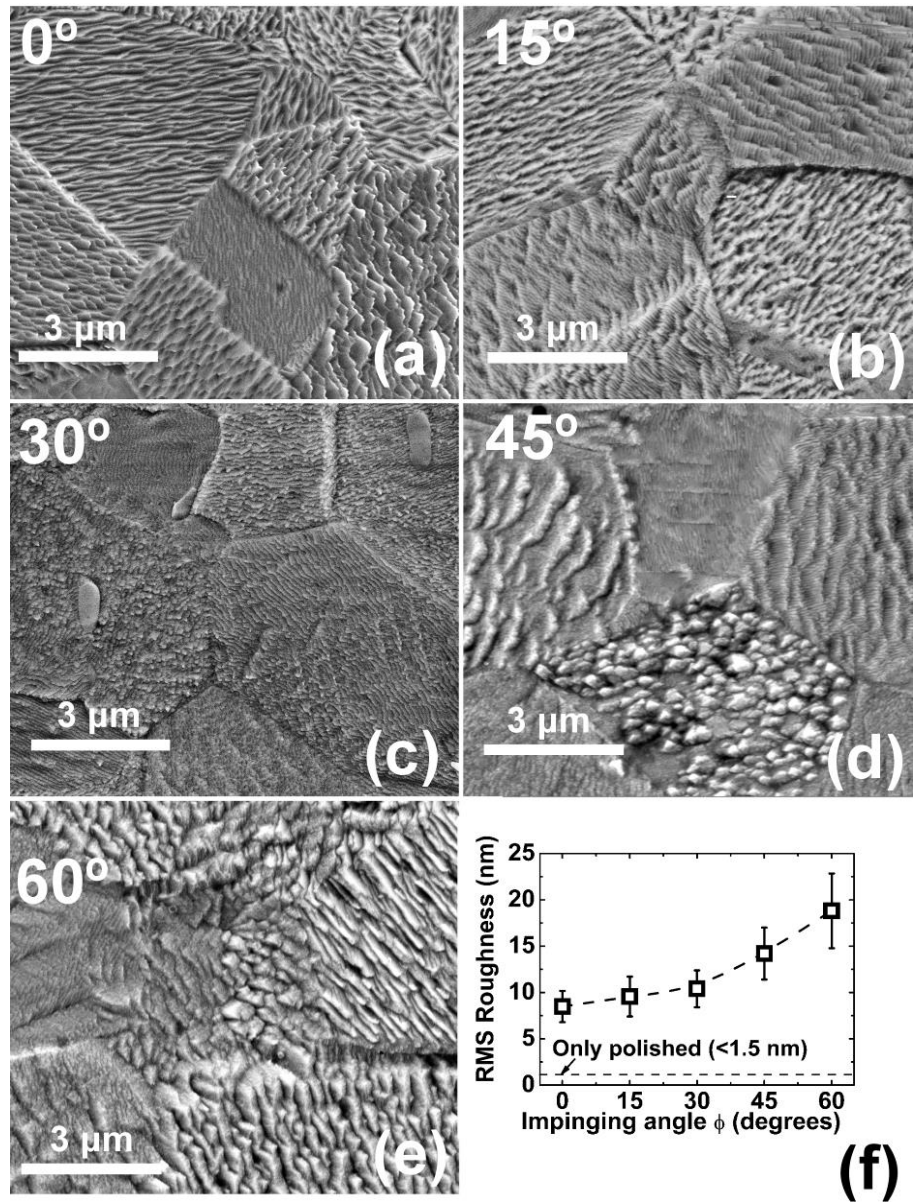

Figure 1: SEM-FEG images from AISI 316L using different ion beam impinging angles φ: (a) 0°; (b) 15°; (c) 30°; (d) 45°; (e) 60° and 1keV fixed energy; (f) RMS roughness measured along several grains (measurements performed with AFM data of area 20μm x 20μm) of the studied AISI 316L samples obtained by bombarding at different impinging angles φ.

Due to the momentum transfer to the sample atoms by the impact of the $Xe^+$ projectiles, impinging angles other than the normal angle modify the formation of patterns. This effect is shown for the studied AISI 316L where, by increasing the ion impinging angles, the roughness of the surface increases (Figure 1(f)). We note that the roughness of the polished samples (RMS roughness <1.5 nm) is much less important than the roughness of the bombarded samples (Figure 1(f)). Besides this, due to the ion bombarding treatment, a sputtering etching (~3 nm/min) takes place, eliminating the surfaces first atomic layers and erasing the effect of the grinding-polished procedure. Going back to the origin of the ripples, the Ehrlich-Schwoebel instability model associates the increasing roughness with larger impinging angles, i.e., the so-called erosion regime [20].

Figure 2(a) shows details obtained by AFM from an AISI 316L sample bombarded at 0º impinging Xe$^+$ ions angle. Figure 2(b) shows a blow-up from the central grain displayed in Figure 2(a). This picture shows a secondary motif along the top of the hill of the ripples, which it is also observed in some grains in Figure 1, confirming the existence of an intricate variety of profiles prompted by the ion bombarding. We suggest that the surface roughness might also contribute to modify the sticking factor of the impinging nitrogen atoms during nitriding processes due to several effects such as multiple collisions [21,22], increasing effective area, and lattice defects acting as a trap for nitrogen on the surface. The probability of multiple collisions for the impinging atoms can be estimated by assuming a triangular (isosceles) roughness ratio giving by $h/2b$. Here, $h$ and $2b$ are the high and the base (width) of the triangular cross-section of the ripples [23]. From the AFM measurements, a ratio $h/2b \cong 0.1$ is estimated for the main ripple patterns as well as for the secondary motif along the top of the hills. Takaishi [23] has shown that the reflected impinging atoms have an average probability of again hitting the surface given by $\alpha = 1 - h/2c$, where $c = (h^2 + b^2)^{1/2}$ is the hypotenuse of the triangle. By substituting the value h/2b=0.1, $\alpha \cong 2\%$. Considering that the same probability could be taken in account along the top of the ripples, an increasing **~4 %** probability of multiple collisions is obtained. Also, AFM measurements show that the roughness generated by the ion bombarding increases the effective area by **~2 %**. All these considerations, together with the increasing surface defects created by the ion bombarding, may contribute to augment the sticking factor of nitrogen in nitriding processes. Figure 3 shows the effect of perpendicular bombarded impact in AISI 4140 sample. The topographic scale shows peaks high up to ~200 nm in samples treated at perpendicular ion impinging angle, also suggesting a larger effective area prompted by the atomic attrition treatment.

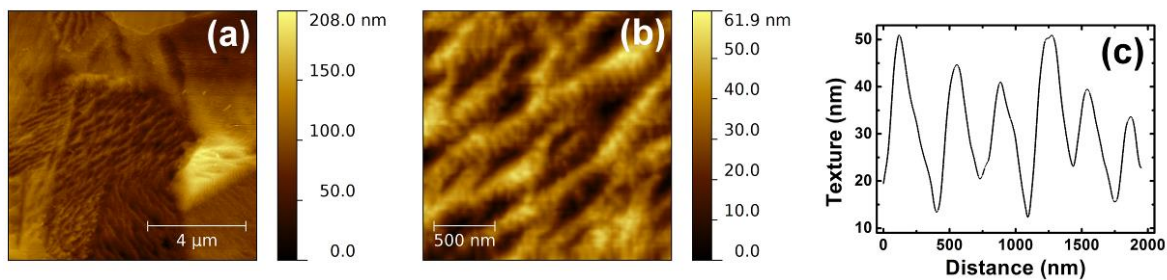

Figure 2: (a) AFM image from AISI 316L perpendicularly bombarded; (b) Detail of the central grain displayed in Figure 2(a); (c) Topographic profile obtained from AFM analysis (for illustrative purposes) along the grain present in Figure 2(b) showing a periodic pattern with average wavelength 0.4 μm.

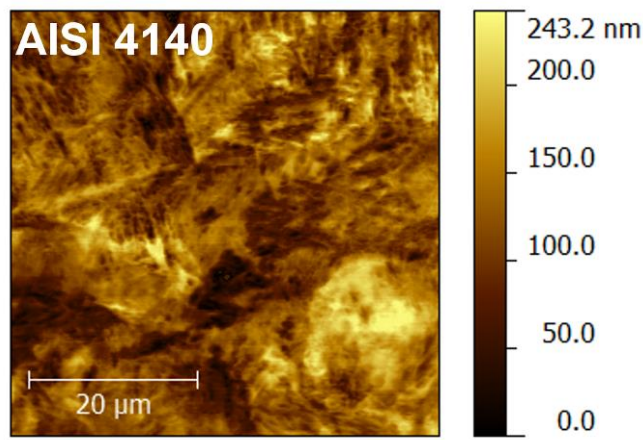

Figure 3: AFM image from AISI 4140 perpendicularly bombarded ($\varphi$ = 0°) showing the surface modifications due to Xe$^+$ ion bombardment.

Although that this paper is not focusing on the study of the ripple formation, we should note that the Ehrlich–Schwoebel instability model discussed above has been recently challenged. Madi *et al.* [24] reported interesting results obtained in p-type doped crystalline silicon bombarded by Ar$^+$ ions claiming that momentum transference and diffusion process are the most important driving forces in the pattern formation. In order to explain the results of Madi *et al.* [25] and Norris *et al.* [26], molecular dynamic (MD) simulation were performed using intrinsic amorphous Si cluster perpendicularly bombarded by Ar$^+$ ions. The theoretical calculations support the Madi´s experimental results and the main conclusion from Norris *et al.* MD calculations is that the erosive effect is not so important on the patterns formation. We note that the experimental differences make difficult a straight forward comparison of our results with those from Madi et al. and Norris et al. [24, 25, 26]. In fact, these authors studied a covalent semiconductor owning stiff bonds and not a metal that is characterized by the non-directional character of its bonds. In metals, on the other hand, due the high diffusivity as compared with semiconductors and to the non-directional character of the metallic bonds there is a tendency to preserve the crystalline structure of the material after bombarding [20]. Therefore, in metals the ripple formation normally follows the crystalline orientation and the Ehrlich–Schwoebel (E–S) model is invoked to explain the experimental findings. We note that other explanations cannot be ruled out but the E–S mechanism seems to be a reasonable approach to the observed results in bombarded metals. Finally, recently much more attention to the stress induced by the noble gases bombarding has been considered as important driving force on the ripple formation [27].

### 3.2. Perpendicular bombardment: effect on the material strain

As discussed above, the effect of perpendicular ion bombardment ($\varphi$ = 0°) on the material strain were investigated in AISI 316L (Section 2.1). As noted in the introduction, the whole phenomenon discussed in this paper are averaged values taken over many grains of the sample. For the study of the in depth average strain profile, the diffractograms for three different X-rays penetration depths were measured by changing the X-ray beam angle impinging the sample (Table 1). Figure 4 shows the peak profiles of the $\gamma(220)$ Bragg reflection for perpendicularly bombarded and non-bombarded samples at the three ion incidence angles.

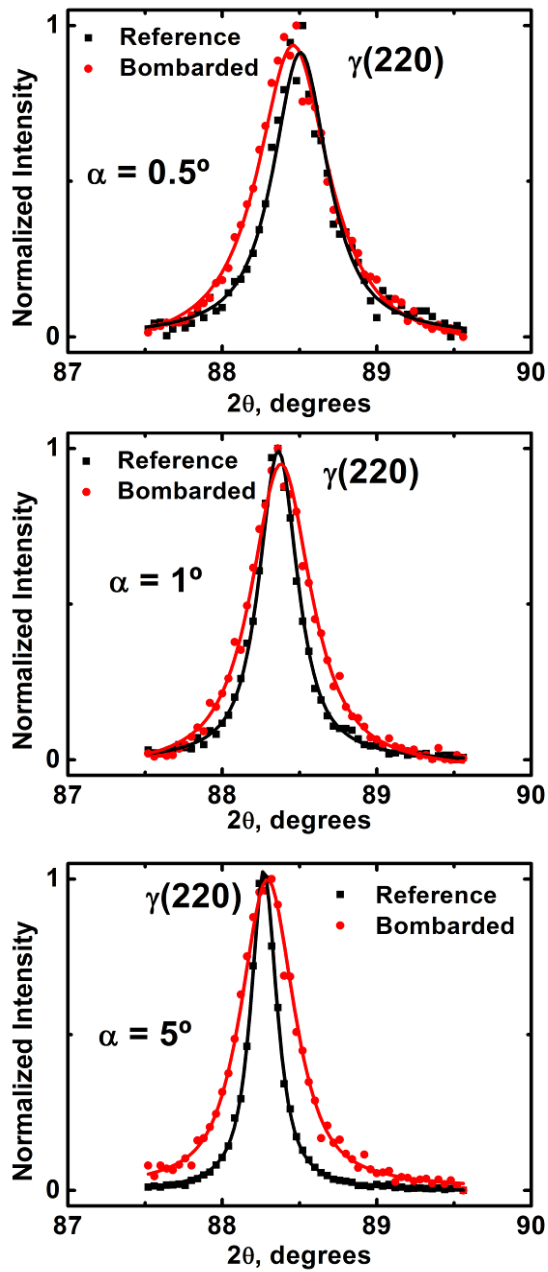

Figure 4: GAXRD diffractograms for selected Bragg reflections (γ-220) from AISI 316L (perpendicularly bombarded and only polished) at different x-ray incident angles.

Figures 5 (a), (b) and (c) show the Williamson-Hall plots obtained from the diffractograms for the studied samples. The Miller indices associated with different reflections are also indicated in the graphs. The experimental error bars are obtained from error-propagation in equation I taking in account the uncertainty introduced by integrating the area of the x-ray peak, as outlined in Section 2.1. The fitting curves shown in Figure 5 (a), (b) and (c) intercept, within the experimental errors, at X~0 and Y~0 preventing us from drawing conclusions about the crystallites size. Therefore, we focus our attention on the curve´s slope, i.e., a magnitude related to the variation of strain generated by ion bombardment. Figure 5 (d) shows the slope ratio for the bombarded ($S_B$) and non-bombarded ($S_{NB}$) samples ($S_B/S_{NB}$) obtained from Figures 5 (a), (b) and (c). No difference between measured strains was noticed in the first 0.2μm of the samples, a region that corresponds to the outmost surface modifications due to preferential sputtering as measured by AFM (Figures 2(a), 2(b) and 3). Nevertheless, there is significant strain difference starting from the region immediately below the top layers modifications due to the bombardment. The evolution of this parameter suggests an augment of the average strain all over the studied x-ray penetration. It is worth noting that the strain generated by the ion bombardment extends orders of magnitude deeper within the sample as compared with the stopping length of the ions (~1.1 – 1.8 nm, Section 2.1).

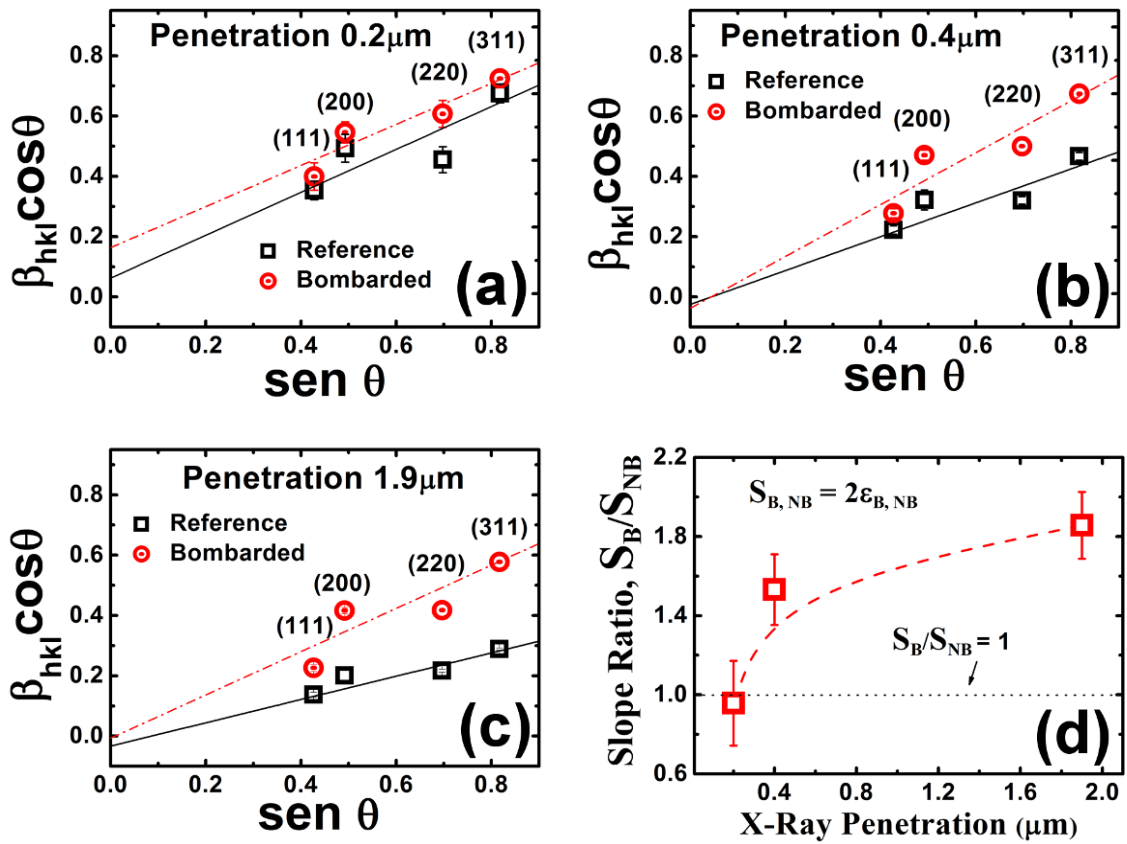

Figure 5: Williamson-Hall curves, $\beta_{hkl}$ cos$\theta$ vs. sin$\theta$, for AISI 316L perpendicularly bombarded ($\varphi$ = 0º) and non-bombarded (only polished) studied depth for: (a) 0.21µm; (b) 0.41µm; (c) 1.89µm (Table I). The Miller indices associated to each reflection are indicated. The slopes are associated to the average strain $\varepsilon$ and the intercept with a mean crystallite size **L;** (d) Slope ratio $S_B/S_{NB}$ for the Bombarded ($S_B$) and Non-Bombarded ($S_{NB}$) samples obtained from plots (a), (b) and (c).

### 3.3. Perpendicular bombarding effect on the nitrided material properties

As explained in Section 2.2, to study the effect of the resultant nanostructured steel surfaces after Xe[+] pre-bombardment on pulsed plasma nitriding process, AISI 316L and AISI 4140 steel samples were specifically prepared. We note that in the experiments reported here, the patterns obtained by masking half of the samples with Si wafers are similar to those obtained *without masking* the substrate. In particular, the micrographs displayed in Figures 1, 2 and 3 were obtained from bombarded samples *without* masking them. This comment is important because, as reported by Zhang *et al.* [28], bombarding Si with Xe[+] ions under simultaneously co-depositing Fe, modify the generated type of pattern. As pointed out in the introduction, the effect of the texture, strain, and defects on the nitriding process studied in this paper are an average phenomenon observed over many grains. Therefore, considering that the nitrogen diffusion dependency on crystal orientation of steel is a well-established result [29], the findings reported in this papers such as diffusion depth and hardness should be considered average values obtained over several grains.

After nitriding, the samples were analyzed by x-ray diffraction to investigate the presence of the nitride compounds. X-rays diffractograms from the analyzed samples (AISI 316L and AISI 4140) show that the phases present in the nitride layers are associated with γ' and the ε phase (Figure 6). In fact, the diffractograms display the phases typically found in standard nitriding processes [30,31]. Moreover, we note that similar diffractograms were found in the nitride layer in samples pre-bombarded and not bombarded, indicating that no new phases are introduced by the pre-bombardment treatment.

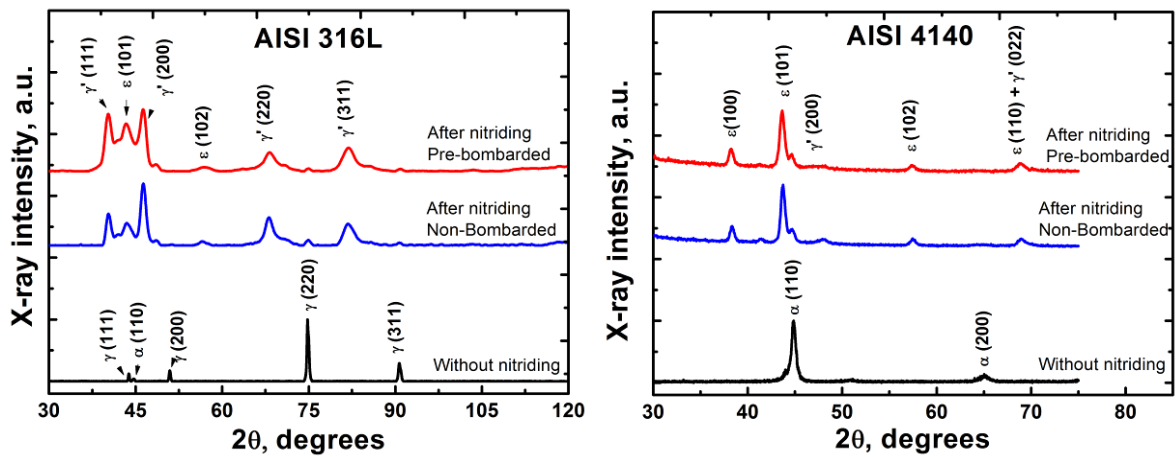

Figure 6: XRD for the studied samples. Left: AISI 316L diffractograms for the pre-bombarded and bombarded samples after nitriding. Right: AISI 4140 diffractograms for the pre-bombarded and bombarded samples after nitriding. The diffractograms for the pristine samples are also displayed.

Figures 7 (bottom) and Figure 8 show the micrograph of the cross-section obtained by SEM in AISI 316L and AISI 4140, respectively, from samples belonging to the sample Set II. The chemical etching reveals the microstructure of the treated samples (see Section 2.2). The characteristic "white" color observed in the micrographs is a finger print of the presence of the iron-nitride phases [32], confirming the results obtained from the diffractograms (Figure 6). The micrographs in Figure 7 show that the total thickness of the nitrides cases is comparable for the two sides of the samples, i.e., the thickness for the nitride case of the non-bombarded (left) and the pre-bombarded (right) parts of the sample are equivalent. Nevertheless, the non-bombarded sample side shows characteristic bi-layers case: one thin layer (~1.5 μm) at the bottom of the layer (indicated by an arrow in Figure 7) followed by a thicker one (~8.6 μm). The presence of a bi-layers case is characteristic of standard non-bombarded nitriding process [30,33]. On the other hand, the micrograph corresponding to the pre-bombarded side of the sample shows a compact nitride monolayer. This remarkable result suggests that the surface modifications introduced by the Xe⁺ bombardment are important to generate a smooth transition between the nitride layer and the diffusion bulk region of the sample. However, future wear experiments will be necessary in order to verify the cohesive quality of this nitride layer.

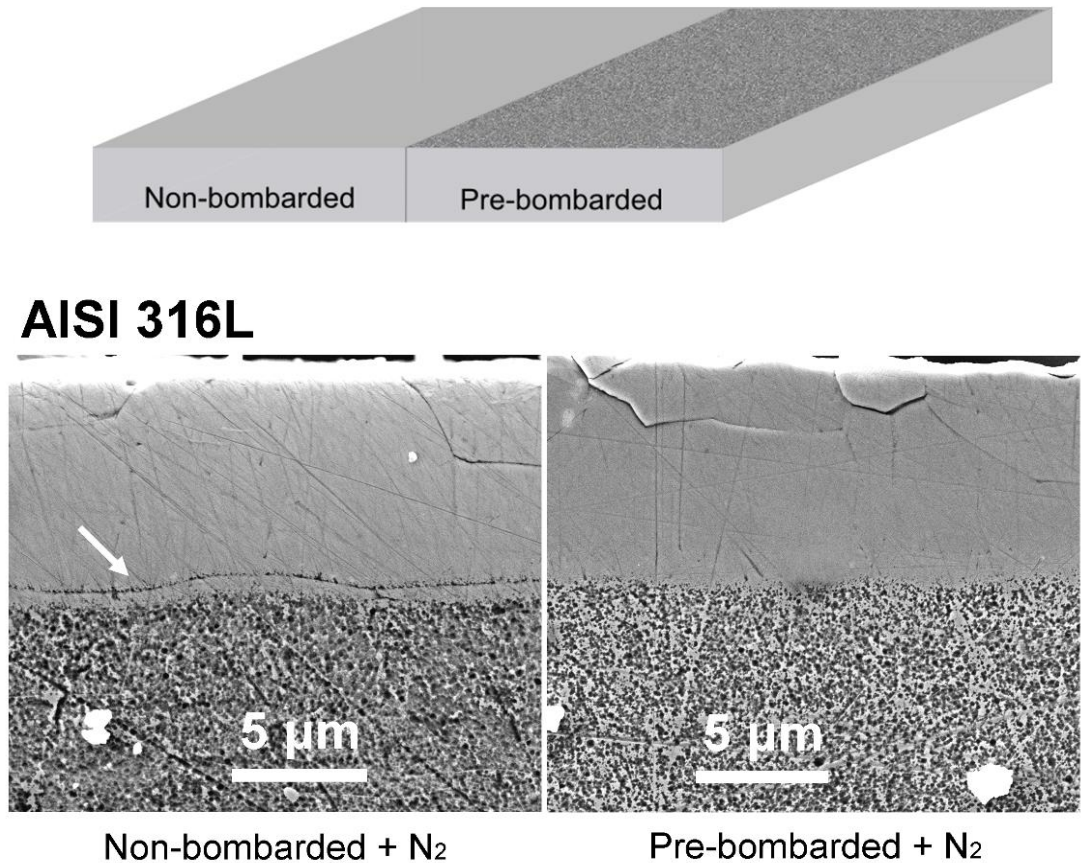

Figure 7: Top: sketch of the cross-section of the studied samples in Set II. The left part of the sample was covered in such a way that only half of it was $Xe^+$ ion perpendicularly bombarded and subsequently nitrided. Bottom: SEM cross-sections micrographs of the AISI 316L nitrided samples. Left (right) sample´s non-bombarded (pre-bombarded) side and afterward pulsed plasma nitrided. The left picture shows the formation of an intermediate thin layer (arrow) absent in the right micrograph.

The nitrided AISI 4140 material also shows peculiarities in both sides of the treated sample (Figure 8). The microstructure of the non-bombarded side shows isolated iron nitride precipitates identified as small white segments' discontinuities (Figure 8, left). On the other hand, the pre-bombarded side shows larger white needle-shaped precipitates (Figure 8, right), indicating enhanced nitrogen diffusion channels. Also, the nitride case of the non-bombarded side is thicker than that of the other side.

# AISI 4140

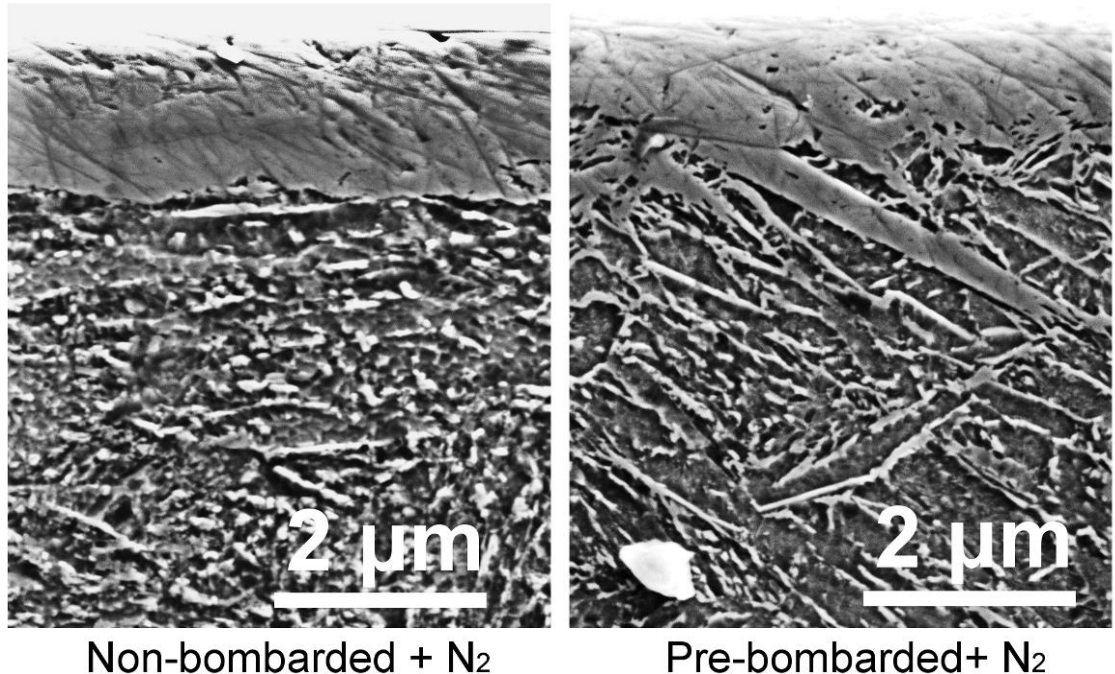

Figure 8: SEM cross-section micrograph of the studied AISI 4140 sample. Left (right) sample´s non-bombarded (pre-bombarded) side and afterward pulsed plasma nitrided. An identical experiment as the indicated in Figure 6, top, was implemented for preparation of the sample before nitriding.

The modifications observed in the microstructure due to the $Xe^+$ ions pre-bombardment also induce changes in the hardness of the material after pulsed plasma nitriding process. Figure 9 (left) shows the hardness profile of the nitrided AISI 316L samples. As observed in the plot, the hardness profile for the pre-bombarded side of the sample is higher than the non-bombarded one in the 10-15 μm region. Also, we note that the hardness gradient is gradual in the pre-bombarded sample as compared with the non-bombarded sample. The more abrupt hardness profile in non-pre bombarded samples probably increases the sharing stress, causing the observed double nitride layer (Figure 7, arrow).

Figure 9 (right) shows the hardness profiles for the AISI 4140 sample. The large statistical error stems from the coarse inhomogeneities (needles) of the nitride layer that introduce large variations in the measurements, i.e., very hard precipitates embedded in relatively soft environment. Nevertheless, the plot shows a trend in the hardness profiles for the pre-bombarded side of the sample.

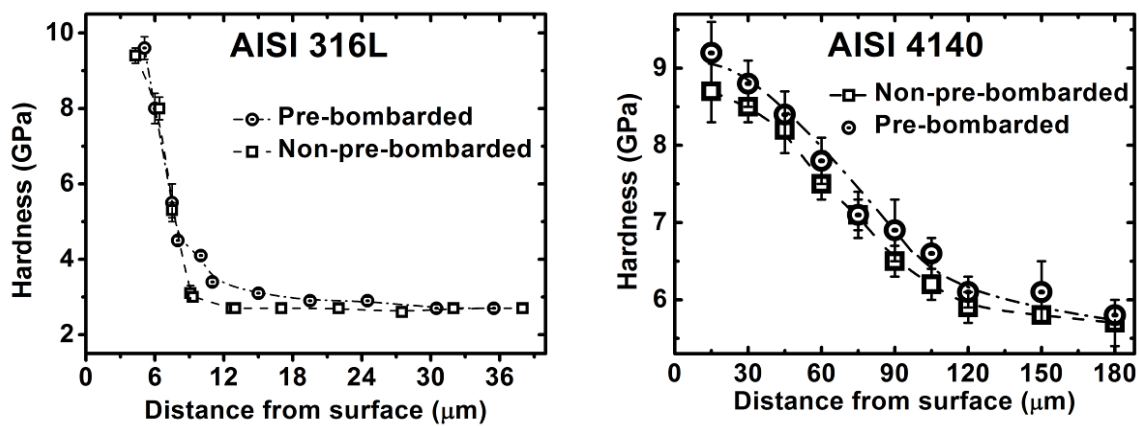

Figure 9: Hardness profile of the nitrided samples. (left) Hardness of AISI 316L as a funcion of the distance from the nitrided surface (the experimental errors are of the size of the symbols).; (right) Hardness of AISI 4140 cross-section as a funcion of the distance from the nitrided surface. The dashed lines are guide for the eyes.

The changes in nitrides' cases for the studied steels are related to the modified surface properties analyzed in sections 3.1 and 3.2. A rougher surface after $Xe^+$ perpendicular bombardment increases nitrogen retention and presence of strain leads to extra diffusion "channels" [34]. Also, bombardment generates defects and stress, enhancing nitrogen diffusion [35].

We note that the bombardment effect on the material properties extends beyond the stopping length of the ions ($\sim 1.1 - 1.8$ nm, Section 2.1). The strain analysis shows clear differences up to approximately 2 $\mu$m depth in perpendicularly bombarded samples, a depth much larger that the stopping lengths estimated by SRIM. The effect of the ion bombarding at distances much deeper than the stopping projectile depth are already reported in the bibliography [7,36,37]. These authors invoke highly anharmonic localized excitations (so called *discrete breath*) propagating distances beyond the ion penetration depth to explain the generation of deeper defects. Also, it is also reported that low energy ion bombarding refines the size of the material grains deeper than the ion penetration [9]. These concomitant modifications could help in understanding the hardness profiles after pulsed plasma nitriding, i.e., prompting differences at distances deeper than the underneath sample regions modified by the ion bombardment. Regarding with the microstructure of nitrides cases, the SEM micrographs show that the formation of the nitride layer as well as the diffusion zone are influenced by the ion bombardment at distances deeper than the ion stopping distance. Some of these effects are probably due to a *knock on* phenomenon generating stressed meta-stable sites and local relaxation by thermal spike [38,39].

## 4. Conclusions

The effect of $Xe^+$ bombardment on the topography, strain, and posterior pulsed plasma nitriding process in AISI 316L and 4140 steel are reported. SEM measurements of $Xe^+$ bombarded samples show the formation of (nanostructured) regular patterns formed on the outmost layers of the crystalline grains of the material. AFM measurements reveal an increasing roughness on the angle of the impinging projectiles. In nitriding process, several concomitant effects are probably contributing to modify nitrogen retention at the material surface such as the increasing roughness, crystalline defects, moderated increasing of effective surface area, and by slightly increasing the atoms collisions probability of re-emitted nitrogen. The perpendicular ion bombardment generates strain at depth orders of magnitude larger than the stopping length of the projectiles, influencing the nitrogen diffusion in the material. SEM measurements reveal that the pre-bombardment modified the iron nitrides microstructure in the pulsed plasma nitriding process. In fact, in pre-bombarded AISI 316L sample, a compact nitride monolayer is obtained without the formation of a thin intermediated layer, as normally found in standard treated samples. The microstructure of the diffusion zone in AISI 4140 samples also depends on the pre-treatment. In pre-bombarded samples, the nitrogen diffusion zone shows long iron nitrides needle-shaped precipitates. This finding is probably associated with diverse nanostructures induced on different grains crystalline orientation, as observed by SEM. On the other hand, in non-bombarded samples, small size iron nitride precipitates are fairly well distributed in the material. The former finding is attributed to diffusion defects channels created by ion bombardment. The hardness profile after pulsed plasma nitriding also depends on the pre-bombardment process. In AISI 316L pre-bombarded samples, higher profile and smoother hardness gradient are observed as compared to the nitrided sample without bombardment. Although a trend is observed in AISI 4140 pre-bombarded samples, an increase in hardness is not conclusive probably due to the presence of coarse hard nitrides precipitates embedded in soft regions of the material introducing a relatively large uncertainty in the nano-hardness measurement.

## Acknowledgments

This work is part of the Master degrees requirements of SC. The authors are grateful to C. Piacenti for technical help. Part of this work was supported by FAPESP, project 2012/10127-5. SC is a FAPESP fellow. FA, LFZ and EAO are CNPq fellows. SEM images were obtained at the Brazilian Nanotechnology National Laboratory (LNNano). The DRX were performed at the Brazilian Synchrotron Light Laboratory (LNLS).